# Stochastic action principle and maximum entropy

Q. A. Wang, F. Tsobnang, S. Bangoup, F. Dzangue, A. Jeatsa and A. Le Méhauté

Institut Supérieur des Matériaux et Mécaniques Avancées du Mans, 44 Av. Bartholdi,
72000 Le Mans, France

**Abstract**

A stochastic action principle for stochastic dynamics is revisited. We present first numerical diffusion experiments showing that the diffusion path probability depend exponentially on average Lagrangian action $A = \int_a^b L dt$. This result is then used to derive an uncertainty measure defined in a way mimicking the heat or entropy in the first law of thermodynamics. It is shown that the path uncertainty (or path entropy) can be measured by the Shannon information and that the maximum entropy principle and the least action principle of classical mechanics can be unified into a concise form $\overline{\delta A}=0$, averaged over all possible paths of stochastic motion. It is argued that this action principle, hence the maximum entropy principle, is simply a consequence of the mechanical equilibrium condition extended to the case of stochastic dynamics.

## 1) Introduction

It is a long time conviction of scientists that the all systems in nature optimize certain mathematical measures in their motion. The search for such quantities has always been a major objective in the efforts to understand the laws of nature. One of these measures is the Lagrangian action considered as a most fundamental quantity in physics. The least action principle[1] [1] has been used to derive almost all the physical laws for regular dynamics (classical mechanics, optics, electricity, relativity, electromagnetism, wave motion, etc.[2]). This achievement explain the efforts to extend the principle to irregular dynamics such as equilibrium thermodynamics[3], irreversible process [4], random dynamics[5][6], stochastic mechanics[7][8], quantum theory[9] and quantum gravity theory[10]. We notice that in most of these approaches, the randomness or the uncertainty (often measured by information or entropy) of the irregular dynamics is not considered in the optimization methods. For example, we often see expression such as $\overline{\delta R} = \delta \overline{R}$ concerning the variation of a random variable $R$ with an expectation $\overline{R}$. This is incorrect because the variation of uncertainty aroused by the variation of the $R$ may play important role in the dynamics.

Another most fundamental measure, called entropy, is frequently used in variational methods of thermodynamics and statistics. The word "entropy" has a well known definition given by Clausius in the equilibrium thermodynamics. But it is also used as a measure of uncertainty in stochastic dynamics. In this sense, it is also referred to as "information" or "informational entropy". In contrast to the action principle, entropy and its optimization have always been a source of controversies. It has been used in different even opposite variational methods based on different physical understanding of the optimization. For instance, there is the principle of maximum thermodynamic entropy in statistical thermodynamics[11][12], the maximum information-entropy[13][14] in information theory, the principle of minimum entropy production [15] for certain nonequilibrium dynamics, and the principle of maximum entropy production for others[16][17]. Certain interpretation of entropy and of its evolution was even thought to be in conflict with the mechanical laws[18]. Notice that these laws can be derived from least action principle. In fact, the definition of entropy is itself a great matter of investigation for both equilibrium and nonequilibrium systems since the proposition of Boltzmann and Gibbs entropy. Concerning the maximum entropy calculus, few people still contest the fact that the maximization of Shannon entropy yields the correct exponential distribution. But curously enough, few people are completely satisfied by the arguments of Jaynes and others[12][13][14] supporting the maximum entropy principle by considering entropy as anthropomorphic quantity and the principle as only an inference method. This question will be revisited to the end of the present paper.

---
[1] We continue to use this term "least action principle" here considering its popularity in the scientific community, although we know nowadays that the term "optimal action" is more suitable because the action of a mechanical system can have a maximum, or a minimum, or a stationary for real paths[19].

In view of the fundamental character of entropy in stochastic dynamics, it seems that the associated variation approaches must be considered as first principles and cannot be derived from other ones (such as least action principle) for regular dynamics where uncertainty does not exist at all. However, a question we asked is whether we can formulate a more general variation principle covering both the optimization of action for regular dynamics and the optimization of information-entropy for stochastic dynamics. We can imagine a mechanical system originally obeying least action principle and then subject to a random perturbation which makes the movement stochastic. For this kind of systems, we have proposed a stochastic action principle [20][21][22] which was originally a combination of maximum entropy principle (MEP) and least action principle on the basis of the following assumptions :

1) A random Hamiltonian system can have different paths between two points in both configuration space and phase space.

2) The paths are characterized uniquely by their action.

3) The path information is measured by Shannon entropy.

4) The path information is maximum for real process.

This is in fact maximization of path entropy under the constraint associated with average action over paths (we assume the existence of this average measure). As expected, this variational principle leads to a path probability depending exponentially on the Lagrangian action of the paths and satisfying the Fokker-Planck equation of normal diffusion[21]. Some diffusion laws such as Fick's laws, Ohm's law, and Fourier's law can be derived from this probability distribution. We noticed that the above combination of two variation principles could be written in a concise form $\overline{\delta A}=0$ [22], i.e., the variation of action averaged over all possible paths must vanish.

However, many disadvantages exist in the above formalism. The first one is that not all the above physical assumptions are obvious and convincing. For example, concerning the path probability, another point of view[23] says that the probability should depend on the average energy on the paths instead of their action. The second disadvantage of that formalism is we used the Shannon entropy as a starting hypothesis, which limits the validity of the formalism. One may think that the principle is probably no more valid if the path uncertainty cannot be measured by the Shannon formula. The third disadvantage is that MEP is already a starting hypothesis, while it was expected that the work might help to understand why entropy goes to maximum.

In this work, the reasoning is totally different even opposite. The only physical assumption we make is a stochastic action principle (SAP), i.e., $\overline{\delta A}=0$. The first and second assumptions mentioned above are not necessary because these properties will be extracted from experimental results. The third and fourth assumptions become purely the consequences of SAP. This work is limited to the classical mechanics of Hamiltonian systems for which the least action principle is well formulated. Neither relativistic nor quantum effects is considered.

## 2) Stochastic dynamics of particle diffusion

We consider a classical Hamiltonian systems moving, maybe randomly, in the configuration space between two points *a* and *b*. Its Hamiltonian is given by $H=T+V$ and its Lagrangian by $L=T-V$ where $T$ is the kinetic energy and $V$ the potential one. The Lagrangian action on a given path is $A=\int_a^b Ldt$ as defined in the Lagrangian mechanics. These definitions need sufficiently smooth dynamics at smallest time scales of observation. In addition, if there are random noises perturbing the motion, the energy variation due to the external perturbation or internal fluctuation is negligible at a time scale $\tau$ which is nevertheless small with respect to the observation period. Hence $L=\overline{T}-\overline{V}$ and $H=\overline{T}+\overline{V}$ can exist, where $\overline{T}$ and $\overline{V}$ are kinetic and potential energies averaged over $\tau$ such as $\overline{T}=\frac{1}{\tau}\int_0^\tau Tdt$.

It is known that if there is no random forces and if the duration of motion $t_{ab}=t_b-t_a$ from *a* to *b* is given, there is only one possible path between *a* and *b*. However, this uniqueness of transport path disappears if the motion is perturbed by random forces. An example is the case of particle diffusion in random media, where many paths between two given points are possible. This effect of noise can be easily demonstrated by a thought experiment in Figure 1. See the caption for detailed description. In this experiment, it is expected that more a path is different from the least action path (straight line in the figure) between *a* and *b*, less there are particles traveling on that path, i.e., smaller is the probability that the path is taken by the particles.

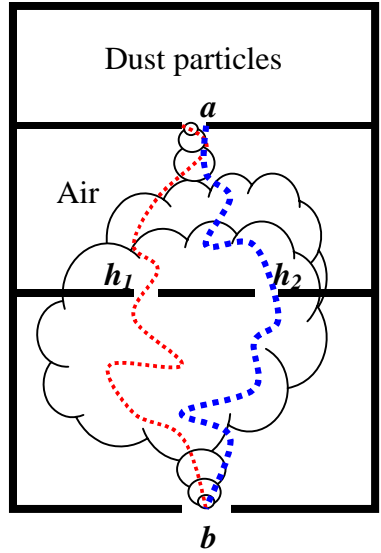

**Figure 1**

A thought experiment for the random diffusion of the dust particles falling in the air. At time $t_a$, the particles fall out of the hole at point *a*. At time $t_b$, certain particles arrive at point *b*. The existence of more than one path of the particles from *a* to *b* can be proved by the following operations. Let us open only one hole $h_1$ on a wall between *a* and *b*, we will observe dust particles at point *b* at time $t_b$. Then close the hole $h_1$ and open another hole $h_2$, we can still observe particles at point *b* at time $t_b$, as illustrated by the two curves passing respectively through $h_1$ and $h_2$. Another observation of this experiment is that more a path is different from the vertical straight line between *a* and *b*, less there are particles traveling on that path, i.e., smaller is the probability that the path is taken by the particles. This observation can be easily verified by the numerical experiment in the following section.

Now let us suppose *W* discrete paths from *a* to *b*. Among a very large *N* particles leaving the point *a*, we observe $N_k$ ones arriving at point *b* by the path *k*. Then the probability for the particles to take the path *k* is defined by $p_{ab}(k) = \frac{N_k}{N}$. The normalization is given by $\sum_k p_{ab}(k) = 1$ or, in the case of continuous space, by the path integral $\int D(r)\, p_{ab} = 1$, where *r* denotes the continuous coordinates of the paths.

## 3) A numerical experiment of particle diffusion and path probability

Does the probability $p_{ab}(k) = \frac{N_k}{N}$ really exist for each path? If it exists, how does it change from path to path? What are the quantities associated with the paths which determines the change in path probability? To answer these questions, we have carried out numerical experiments (Figure 2) showing the dust particles fall from a small hole *a* on the top of a two dimensional experimental box to the bottom of the box. A noise is introduced to perturb symmetrically in the direction of *x* the falling particles. We have used three kind of noises: Gaussian noise, uniform noise (with amplitudes uniformly distributed between -1 and 1) and truncated uniform noises (uniform noise with a cutoff of magnitude between -*z* and *z* where $z<1$, i.e., the probability is zero for the magnitude between –z and z).

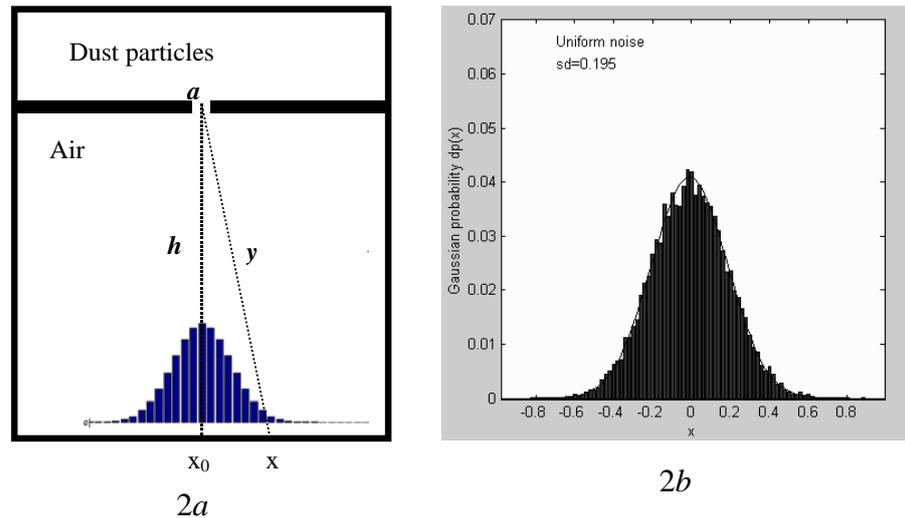

**Figure 2**

**2a**: Model of the numerical experiment showing the dust particles fall from a small hole *a* onto the bottom of the experimental box. The distribution of particles on the bottom (represented by the vertical bars) is caused by the random noise (air for example) in the direction of *x*. **2b**: An example of experimental results in which the falling particles are perturbed by a noise whose magnitude is uniformly distributed between -1 and 1 in *x*. The vertical bars are experimental result and the curve is a Gaussian distribution $dp(x) = \frac{dN(x)}{N} = \frac{1}{\sigma\sqrt{2\pi}} \exp(-\frac{(x-x_0)^2}{2\sigma^2})dx$, where *dN(x)* is the particle number in the interval *x—x+dx*, *N* is the total number of falling particles and σ is the standard deviation (*sd*). The experiments show that the *dp(x)* is always Gaussian whatever the noise (uniform, Gaussian or other truncated uniform noises).

The observed distributions of particles are Gaussian for the three noises. The standard deviation of the distributions is uniquely determined by the nature of the noise (type, maximal magnitude, frequency etc.). This result was expected because of the finite variances of the used noises and of the central limit theorem saying that the attractor distribution is a normal (Gaussian) one if the noises (random variables) have finite variance.

What can we conclude from this experiment of falling particles which seems to be trivial?

First, let us suppose that the falling distance $h$ is small so that the path $y$ between $a$ and any position $x$ on the bottom can be considered as a straight line and the average velocity on $y$ can be given by $\frac{y}{\tau}$ where $\tau$ is the motion time from $a$ to $x$ (see Figure 2a). In this case, it is easy to show that the action $A_x$ from $a$ to $x$ is proportional to $(x-x_0)^2$, i.e.,

$$A_x = (\bar{T} - \bar{V})\tau = m\frac{y^2}{2\tau} - \frac{mgh}{2}\tau = m\frac{y^2}{2\tau} - \frac{mgh\tau^2}{2\tau} = m\frac{y^2}{2\tau} - \frac{mh^2}{\tau} = \frac{m}{2\tau}(x-x_0)^2 - \frac{mh^2}{2\tau}$$

where $\bar{T} = m\frac{y^2}{2\tau}$ and $\bar{V} = \frac{mgh}{2}$ are the average kinetic and potential energy, respectively. This analysis applies to any smooth motion provided $h$ is small. Considering the observed Gaussian distribution of the falling particles in figure 2, we can write for small $h$ $dN(x) \propto \exp(-\eta A_x)$ where $\eta$ is a constant. The probability that a particle takes the small straight path from $a$ to $x$ is proportional to the exponential of action $A_x$.

Now let us consider large $h$. In this case the paths may not be straight lines. But a curved path from $a$ to $x$ can be cut into small intervals at $x_1, x_2, \ldots$. The above analysis is still valid for each small segment. The probability that a particle takes the path to $x$ is then equal to the product of the probabilities on every segment of that path from $a$ to $x$ and should be proportional to the exponential of the total action from $a$ to $x$, i.e.,

$$p_{ax} \propto \prod_i \exp(-\eta A_i) = \exp(-\eta \sum_i A_i) = \exp(-\eta A_{ax}) \tag{1}$$

where $A_i$ is the action on the segment $x_i$ and $A_{ax}$ is the total action on a given path from $a$ to $x$. The constant $\eta$ is a characteristic of the noise and should be the same for every segment. The conclusion of this section is the path probability depends exponentially on action *as long as the particle distribution on the bottom of the box is Gaussian.*

Concerning the exponential form of path probability, there is another proposal [23] $p_{ab}(k) \propto \exp(-\gamma \overline{H_k})$, i.e., the path probability depends exponentially on the negative average energy. According to this probability, the most probable path has minimum average energy, so that for vanishing noise (regular dynamics), this minimum energy path would be the unique one which must also follow the least action principle. Here we have a paradox because the real path given by least action principle is in general not the path of minimum average energy.

## 4) An action principle for stochastic dynamics

Recently, the following stochastic action principle (SAP) was postulated[20][22] :

$$\overline{\delta A} = 0 \tag{2}$$

where $\overline{\delta A} = \int D(r) p_{ab} \delta A$ is the average of the variation $\delta A$ over all the paths. It can be written as follows

$$\begin{aligned} \overline{\delta A} &= \int D(r) p_{ab} \delta A \\ &= \delta \int D(r) p_{ab} A - \int D(r) A \delta p_{ab} \\ &= \delta \overline{A} - \frac{1}{\eta} \delta S_{ab} \end{aligned} \tag{3}$$

where $\overline{A} = \int D(r) p_{ab} A$ is the ensemble average of action $A$, and $\delta S_{ab}$ is defined by

$$\delta S_{ab} = \eta \left( \delta \overline{A} - \overline{\delta A} \right) = \eta \int D(r) A \delta p_{ab}. \tag{4}$$

Eq.(4) makes it possible to derive $S_{ab}$ directly from probability distribution if the latter is known. Let us consider the dynamics in the section 3 that has the *exponential path probability*

$$p_{ab} = \frac{1}{Z} \exp(-\eta A) \tag{5}$$

where $Z = \int D(r) \exp(-\eta A)$ is the partition function of the distribution. A trivial calculation tell us that $\delta S_{ab}$ is a variation of the path entropy $S_{ab}$ given by Shannon formula

$$S_{ab} = -\int D(r) p_{ab} \ln p_{ab}. \tag{6}$$

Eq.(4) is a definition of entropy or information as a measure of uncertainty of random variable (action in the present case)[26]. It mimics the first law of thermodynamics $dQ = dU + dW$ where $U = \overline{E} = \sum_i p_i E_i$ is the average energy, $E_i$ is the energy of the state $i$ with probability $p_i$, $dW$ is the work of the forces $F_j = -(\sum_i p_i \frac{\partial E_i}{\partial q_j})$ and $q_j$ is some extensive variables such as volume, surface, magnetic moment etc. The work can be written as $dW = -\sum_j \left( \sum_i p_i \frac{\partial E_i}{\partial q_j} \right) dq_j = -\sum_i p_i dE_i = -\overline{dE_i}$. So the first law becomes $dQ = d\overline{E} - \overline{dE}$. We see that by Eq.(4) a "heat" $Q$ is defined as the measure the randomness of action (or of any other random variables in general[26]). In Eq.(6), this heat" is related to the Shannon entropy since the probability is exponential. If the probability is not exponential, the functional of the entropy is probably different from the Shannon one, as discussed in [26].

With the help of Eqs.(2) and (5), it is easy to verify that

$$\delta p_{ab} = -\eta p_{ab} \delta A \qquad (7)$$

and

$$\delta^2 p_{ab} = -\eta p_{ab} \delta^2 A. \qquad (8)$$

From Eqs.(7) and (8), the maximum condition of $p_{ab}$, i.e., $\delta p_{ab} = 0$ and $\delta^2 p_{ab} < 0$, is transformed into $\delta A = 0$ and $\delta^2 A > 0$ if the constant $\eta$ is positive, that is *the least action path is the most probable path*. On the contrary, if $\eta$ is negative, we get $\delta A = 0$ and $\delta^2 A < 0$, the *most probable path is a maximum action one*.

In our previous work, we have proved that the probability distribution of Eq.(5) satisfied the Fokker-Planck equation in configuration space. It is easy to see that[20], in the case of free particle, Eq.(5) gives us the transition probability of Brownian motion with $\eta = \frac{1}{2mD} > 0$ where $m$ is the mass and $D$ the diffusion constant of the Brownian particle[25].

## 5) Return to the regular least action principle

The stochastic action principle Eq.(2) should recover the usual least action principle $\delta A = 0$ when the stochastic dynamics tends to regular dynamics with vanishing noise. To show this, let us put the probability Eq.(5) into Eq.(6), a straightforward calculation leads to

$$S_{ab} = \ln Z + \eta \overline{A}. \qquad (9)$$

In regular dynamics, $p_{ab} = 1$ for the path of optimal (maximal or minimal or stationary) action $A_0$ and $p_{ab} = 0$ for other paths having different actions, so that $S_{ab} = 0$ from Eq.(6). We have only one path, the integral in the partition function gives $Z=\int D(q)\exp(-\eta A)=\exp(-\eta A_0)$. Eq.(9) yields $\overline{A}=A_0$. On the other hand, we have $\delta S_{ab}=\eta(\delta\overline{A}-\overline{\delta A})=0$. Thus our principle $\overline{\delta A}=0$ implies $\delta\overline{A}=\delta A_0=0$ or, more generally, $\delta A=0$. This is the usual action principle.

## 6) Stochastic action principle and maximum entropy

Eq.(3) tells us that the SAP given by Eq.(2) implies

$$\delta(S_{ab}-\eta\overline{A})=0. \qquad (10)$$

meaning that the quantity $(S_{ab}-\gamma\overline{A})$ should be optimized. If we add the normalization condition, the SAP becomes:

$$\delta[S_{ab}-\eta\overline{A}+\alpha(\int D(q)p_{ab}-1)]=0 \qquad (11)$$

which is just the usual Jaynes principle of maximum entropy. Hence Eq.(2) is equivalent to the Jaynes principle applied t path entropy.

Is Eq.(2) simply a concise mathematical form of Jaynes principle associated to average action? Or is there something of fundamental which may help us to understand why entropy gets to maximum for stable or stationary distribution?

From section 4, we understand that, in the case of equilibrium system, the variation $\overline{dE_i}$ is a work $dW$. However, in the case of regular mechanics, $dW=0$ is the condition of equilibrium meaning that the sum of all the forces acting on the system should be zero and the net torque taken with respect to any axis must also vanish. So it seems reasonable to take $\overline{dE_i}=0$ as an equilibrium condition for stochastic equilibrium. In other words, when a random motion is in (global) equilibrium, the total work $dW=-\sum_j\left(\sum_i p_i\frac{\partial E_i}{\partial q_j}\right)dq_j$ by all the random forces $f_j=\frac{\partial E_i}{\partial q_j}$ on all the virtual increments $dq_j$ of a state variable (e.g., volume) must vanish. As a consequence of the first law, $\overline{dE_i}=0$ naturally leads to $\delta[S-\eta U]=0$, i.e., Jaynes maximum entropy principle associated with the average energy $\delta[S-\eta U+\alpha 1]=0$ where $S$ is the thermodynamic entropy. This analysis seems to say that the maximum entropy (maximum randomness) is required by the mechanical equilibrium condition in stochastic situation. Remember that $\overline{dE_i}$ can also be written as a variation of free energy $F=U-TS$, i.e., $\overline{dE_i}=dF$. The stochastic equilibrium condition can be put into $dF=0$.

Coming back to our SAP in Eq.(2), the system is in nonequilibrium motion. If there is no noise, the true path satisfies $\delta A=0$ and $\frac{\partial}{\partial t}\frac{\partial L}{\partial \dot{r}_j}-\frac{\partial L}{\partial r_j}=0$. When there is noise perturbation, we have[22]

$$\overline{dA}=\sum_j\left[\int D(r_j)p_{ab}\int_a^b\left(\frac{\partial}{\partial t}\frac{\partial L}{\partial \dot{r}_j}-\frac{\partial L}{\partial r_j}\right)dt\right]dr_j=0 \qquad (12)$$

where $\frac{\partial}{\partial t}\frac{\partial L}{\partial \dot{r}_j}-\frac{\partial L}{\partial r_j}=f_j\neq 0$ is the random force on $dr_j$. Let $\overline{f_j}=\frac{1}{t_{ab}}\int_a^b f_j dt$ be the time average of the random force $f_j$, we obtain

$$\overline{dA}=t_{ab}\sum_j\left[\int D(r_j)p_{ab}\overline{f_j}\right]dr_j=t_{ab}\overline{\overline{dW}}=0 \qquad (13)$$

where $\overline{\overline{dW}}=\sum_j\left[\int D(r_j)p_{ab}\overline{f_j}\right]dr_j=\sum_j\int D(r_j)p_{ab}\overline{dW_j}$ is the ensemble average (over all paths) of the time average $\overline{dW_j}=\frac{1}{t_{ab}}\int_a^b f_j dr_j dt=\frac{1}{t_{ab}}\int_a^b dW_j dt$ and $dW_j=f_j dr_j$ is the work of random force over the variation (deformation) $dr_j$ of a given path. Eq.(13) means

$$\overline{\overline{dW}}=0 \qquad (14)$$

since $t_{ab}$ is arbitrary. Eq.(14) implies that the average work of the random forces at any moment over any time interval and over arbitrary path deformation must vanish. This condition can be satisfied only when the motion is totally random, a state at which the system does not have privileged degrees of freedom without constraints. Indeed, it is easy to show that the maximum entropy with only the normalization as constraint yields totally equiprobable paths. This argument also holds for equilibrium systems. The vanishing work $\overline{dE_i}=dW=0$ needs that, if there is no other constraint than the normalization, no degree of freedom is privileged, i.e., all microstates of the equilibrium state should be equiprobable. This is the state which has the maximum randomness and uncertainty.

To summarize this section, the optimization of both equilibrium entropy and nonequilibrium path entropy is simply the requirement of the mechanical equilibrium conditions in the case of stochastic motion. There is no mystery in that. Entropy or dynamical randomness (uncertainty) must take the largest value for the system to reach a state where the total virtual work of the random forces should vanish. Entropy is not necessarily anthropomorphic quantity as claimed by Jaynes[14] to be able to take maximum for correct inference. Entropy is nothing but a measure of physical uncertainty of stochastic situation. Hence maximum entropy is not merely an inference principle. It is a law of physics. This is a major result of the present work.

## 7) Concluding remarks

We have presented numerical experiments showing the path probability distribution of *some* stochastic dynamics depends exponentially on Lagrangian action. On this basis, a stochastic action principle (SAP) formulated for Hamiltonian system perturbed by random forces is revisited. By using a new definition of statistical uncertainty measure which mimics the heat in the first law of equilibrium thermodynamics, it is shown that, *if the path probability is exponential of action*, the measure of path uncertainty we defined is just Shannon information entropy. It is also shown that the SAP yields both the Jaynes principle of maximum entropy and the conventional least action principle for vanishing noise. It is argued that the maximum entropy is the requirement of the conventional mechanical equilibrium condition for the motion of random systems to be stabilized, which means the total virtual work of random forces should vanish at any moment within any arbitrary time interval. This implies, in equilibrium case, $\overline{dE_i}=0$, and in nonequilibrium case, $\overline{dA} = \overline{\overline{dW}} = 0$. In both cases, the randomness of the motion must be at maximum in order that all degrees of freedom are equally probable if there is no constraint. By these arguments, we try to give the maximum entropy principle, considered by many as only an inference principle, the status of a fundamental physical law.